\documentclass[prl,aps,twocolumn,groupedaddress,showpacs,floatfix]{revtex4}
\usepackage{amsmath,amssymb,multirow,epsfig,bm}

\newcommand{\etal}{{\it et al.,\;}}
\newcommand{\beq}{\begin{equation}}
\newcommand{\eeq}{\end{equation}}
\newcommand{\bea}{\begin{eqnarray}}
\newcommand{\eea}{\end{eqnarray}}

\newcommand{\tr}{\mathrm{Tr}}

\newcommand{\nn}{\nonumber}
\newcommand{\benn}{\begin{displaymath}}
\newcommand{\eenn}{\end{displaymath}}

\begin{document}
\title{The Finite Temperature Pairing Gap of a Unitary Fermi Gas by
\mbox{Quantum Monte Carlo Calculations}}

\author{Piotr Magierski$^1$, Gabriel Wlaz\l{}owski$^1$, Aurel Bulgac$^2$,
and Joaqu\'{\i}n E. Drut$^{2*}$ }

\affiliation{$^1$Faculty of Physics, Warsaw University of Technology,
ulica Koszykowa 75, 00-662 Warsaw, POLAND }
\affiliation{$^2$Department of Physics, University of Washington, Seattle,
WA 98195--1560, USA}

\begin{abstract}
We calculate the one-body temperature Green's (Matsubara) function of the
unitary Fermi gas via Quantum Monte Carlo, and extract the spectral weight
function $A(p,\omega)$ using the methods of maximum entropy and singular
value decomposition. From $A(p,\omega)$ we determine the quasiparticle
spectrum, which can be accurately parametrized by three functions of
temperature:
an effective mass $m^*$, a mean-field potential $U$, and a gap $\Delta$.
Below the critical temperature $T_c=0.15\varepsilon_F$ the results for
$m^*$, $U$ and $\Delta$ can be accurately reproduced using an independent
quasiparticle model. We find evidence of a pseudogap in the fermionic
excitation spectrum for temperatures up to \mbox{$T^*\approx
0.20\varepsilon_{F} > T_c$}.
\end{abstract}

\date{\today}


\maketitle

Over the last few years, the field of cold atoms has drawn unprecedented attention,
as documented in two recent review articles \cite{reviews}. Within this field, the
unitary Fermi gas, defined as the limit of vanishing interaction range and infinite
scattering length, continues to be a fascinating area of research for a number of reasons.
First, the properties of the unitary regime are universal, making this problem relevant
to a wide range of fields including string theories, the quark-gluon plasma,
neutron stars, nuclei, and to a certain extent to
high $T_c$-superconductors. Secondly, experimentalists can control the
strength of the interaction by means of Feshbach resonances, which
allows for the systematic exploration of weakly as well as strongly
coupled regimes.
Finally, these systems exhibit a rich variety of phenomena and properties
(many of which await verification), creating an ideal playground for a large
set of many-body techniques, possibly the largest ever applied to a single problem.

Properties established so far include: energy as a function of
temperature, entropy, frequencies of collective modes, speed of sound,
critical temperature for the onset of superfluidity, and moment of
inertia.
In the case of polarized Fermi systems, the critical spin polarization at
which superfluidity disappears has also been determined.
All of these properties have been established quantitatively with a
reasonable
degree of certainty and accuracy, both experimentally and theoretically.

In spite of great efforts on the part of both theorists and
experimentalists, some fundamental properties of these systems remain 
unknown. Among the most pressing
questions is the magnitude of the pairing gap and its evolution with
temperature.
Theoretical progress in this direction has been nearly at a standstill,
except for the theoretical
determination of the pairing gap at $T=0$ in
Refs.\cite{carlson,chang,cr,gc}, the
recent analysis of experimental data of Ref.\cite{reddy} and a recent
experiment\cite{andre}.

While theoretical models abound, predictions are mostly qualitative,
and their validity and accuracy are difficult to assess due to the
absence of a small parameter for a Fermi gas at unitarity. This work presents the first
{\it ab initio} evaluation of the one-body temperature propagator of the
unitary Fermi gas, free of uncontrolled approximations, which allows for the
extraction of the temperature dependence of the pairing (pseudo)gap \cite{huefner}.

We begin by defining the one-body temperature Green's (Matsubara) function
\cite{fw}:
\bea
{\cal G}(\bm{p},\tau)=
\frac{1}{Z} \tr \{\exp[-(\beta-\tau) (H-\mu N)]\psi^\dagger(\bm{p}) \times
\nn \\
\exp[-\tau(H-\mu N)\psi(\bm{p})] \}, \label{eq:Gp}
\eea
where $\beta = 1/T$ is the inverse temperature and $\tau >0$.
The trace $\tr$ is
performed over the Fock space, and $Z=\tr \{\exp[-\beta
(H-\mu N)]\}$. The spectral weight function $A(\bm{p},\omega)$
can be extracted from the temperature Green's function using the relation:
\beq
{\cal G }(\bm{p},\tau)=-\frac{1}{2\pi}\int_{-\infty}^{\infty}
d\omega A(\bm{p},\omega)\frac{\exp(-\omega\tau)}{1+\exp(-\omega\beta)}.
\label{eq:Ap}
\eeq
By definition, $A(\bm{p},\omega)$ fulfills the following constraints:
\beq
A(\bm{p},\omega) \ge 0, \quad \quad
\int_{-\infty}^{\infty}\frac{d\omega}{2\pi} A(\bm{p},\omega) = 1
\label{eq:Ap_con}.
\eeq
Since our study focuses on the spin-symmetric system, and the Hamiltonian is
spin-symmetric as well, ${\cal G }(\bm{p},\tau)$ is diagonal in the spin
variables and these are suppressed in all formulas.
The numerical evaluation of the one-body temperature propagator
(\ref{eq:Gp}) is performed as
described in Refs.\cite{bdm,bcs-bec}, by using a Trotter expansion of
$\exp[-\tau(H-\mu N)]$,
followed by a Hubbard-Stratonovich transformation of the interaction and an
evaluation of the emerging path-integral via Metropolis importance
sampling. The number of imaginary time steps required to obtain an
accuracy smaller than the statistical error varies with temperature. At low
temperatures the number of time steps is ${\cal{O}}(10^3)$ \cite{bcs-bec}.
All
calculations presented here have been performed with an average total
particle number of 50-55 on an $8^3$ lattice with periodic boundary
conditions \cite{bdm}. We have generated between 6000
and 10000 uncorrelated samples at each temperature and the statistical
errors are typically below 1\%. The systematic errors, some due to
finite lattice effects,
others due to finite range effects,
are estimated at about 10-15\%. Our $T=0$ extrapolation results
\cite{bcs-bec} for the energy per particle are systematically lower
than previous fixed node Monte Carlo results which are variational
\cite{carlson,chang,giorgini}. We have not used the fixed-node
approximation and the value for $\xi=5E/3N\varepsilon_F\approx 0.40$ that we
extract
at unitarity is in agreement with the auxiliary field Monte Carlo results
of Ref. \cite{new-mc}.

The numerical determination of $A(\bm{p},\omega)$ via inversion of
Eq.~(\ref{eq:Ap})
is an ill-posed problem that requires special methods. We have used two,
based on completely different approaches. The first approach is the maximum entropy
method
\cite{jaynes}, which is based on Bayes' theorem.
Quantum Monte Carlo (QMC) calculations provide us with a discrete set of
values
${\cal \tilde{G}}(\bm{p},\tau_{i})$, where $i=1,2,...,{\cal
N}_{\tau}=50$. We treat them as normally distributed random numbers around
the true values
${\cal G}(\bm{p},\tau_{i})$.
The Bayesian strategy consists in maximizing the {\em posterior
probability}
$P(A|\tilde{G}) \propto P(\tilde{G}|A)P(A)$ of finding the right
$A(\bm{p},\omega)$ under
the condition that ${\cal \tilde{G}}(\bm{p},\tau_{i})$ are known. Here,
$P(\tilde{G}|A)\propto\exp\left (-\frac{1}{2}\chi^{2}\right )$ is the {\em
likelihood
function}, where $\chi^{2}=\sum_{i=1}^{{\cal N}_{\tau}}
\left [{\cal \tilde{G}}(\bm{p},\tau_{i}) - {\cal G}(\bm{p},\tau_{i})
\right]^{2}/ \sigma^{2} .$
The quantity ${\cal G}(\bm{p},\tau_{i})$ is determined by the spectral
weight function in the discretized form of Eq.~(\ref{eq:Ap}) at
frequencies $\omega_k$.
The prior probability $P(A)$, describing our ignorance about the spectral
weight function, is defined as $P(A)\propto\exp(\alpha S({\cal M}))$,
where
$\alpha>0$ and $S({\cal M})$ is the relative information entropy with
respect to
the assumed model ${\cal M}$:
\bea
S({\cal M})&=& \sum_{k}\Delta\omega \biggl [ A(\bm{p},\omega_{k}) - {\cal
M}(\omega_{k})
\nn \\
& -& A(\bm{p},\omega_{k})\ln
\left ( \frac{A(\bm{p},\omega_{k})}{{\cal M}(\omega_{k})} \right ) \biggr ]
. \label{eq:method}
\eea
Hence the maximization of $P(A|\tilde{G})$ leads in practice to the
minimization
of the quantity $\frac{1}{2}\chi^{2} - \alpha S({\cal M})$ with respect to
$A$. Note that the
parameter $\alpha$ governs the relative importance of the two terms. The
entropy term prevents excessive inclusion of unjustified structure into
the shape of the spectral weight function. The constraints (\ref{eq:Ap_con})
are enforced by means of Lagrange multipliers.

The second approach is based on the singular value decomposition of
integral kernel ${\cal K}$ of Eq.~(\ref{eq:Ap}), which can be rewritten in
operator form as
\beq
{\cal G}(\bm{p},\tau_{i})=({\cal K}{A})(\bm{p},\tau_{i}).
\eeq
The operator ${\cal K}$ possesses a singular system defined as:
\beq
{\cal K}{u}_{i}=\lambda_{i}\vec{v}_{i}, \quad
{\cal K}^{*}\vec{v}_{i}=\lambda_{i}{u}_{i},
\eeq
where ${\cal K^{*}}$ denotes the adjoint of ${\cal K}$, the $\lambda_{i}$
are the singular
values and the ${u}_{i}$, $\vec{v}_{i}$ are right-singular functions and
left-singular
vectors
respectively. The singular system forms a suitable basis for the expansion
of the spectral weight
function \cite{svd1}, which we can then write as
\beq
{A}(\bm{p},\omega)=\sum_{i=1}^{r}b_{i}(\bm{p}){u}_{i}(\omega),
\label{A:exp}  \quad
b_{i}(\bm{p})=\frac{1}{\lambda_{i}}( \vec{\cal G}(\bm{p})\cdot \vec{v}_{i}
),
\eeq
where $[\ \cdot \ ]$ is a scalar product and $r$ is the rank of the operator
${\cal K}{\cal K^{*}}$. Since ${\cal G}(\bm{p},\tau_{i})$ is affected by the QMC
errors $\sigma_i$, the coefficients $b_{i}$ carry some uncertainty $\Delta
b_{i}$. Each set of expansion coefficients \mbox{$\tilde{b}_{i}\in (b_{i}-\Delta
b_{i},b_{i}+\Delta b_{i})$} reproduces ${\cal G}(\bm{p},\tau_{i})$ within its
error
bars. We use this flexibility of choosing the expansion coefficients to
produce a solution satisfying constraints (\ref{eq:Ap_con}) \cite{svd2}.

\begin{figure}[htb]
\includegraphics[width=7.5cm]{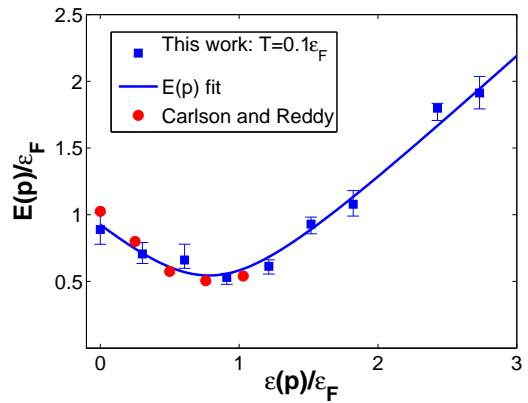}
\caption{ (Color online)
Quasiparticle energies $E({\bm p})$ (squares) extracted
from the spectral weight function $A(\bm{p},\omega)$ at
$T=0.1\varepsilon_{F}$.
The line corresponds to the fit
to Eq.~(\ref{eq:ek}). The circles are the results of Carlson and Reddy
\cite{cr}.
\label{fig:eqp} }
\end{figure}

The advantages and disadvantages of both methods will be discussed
elsewhere \cite{mw}. Here we note only that since they are based on
completely different
approaches their agreement serves as a robust test for the determination of
the spectral weight function.
\begin{figure}[htb]
\includegraphics[width=7.5cm]{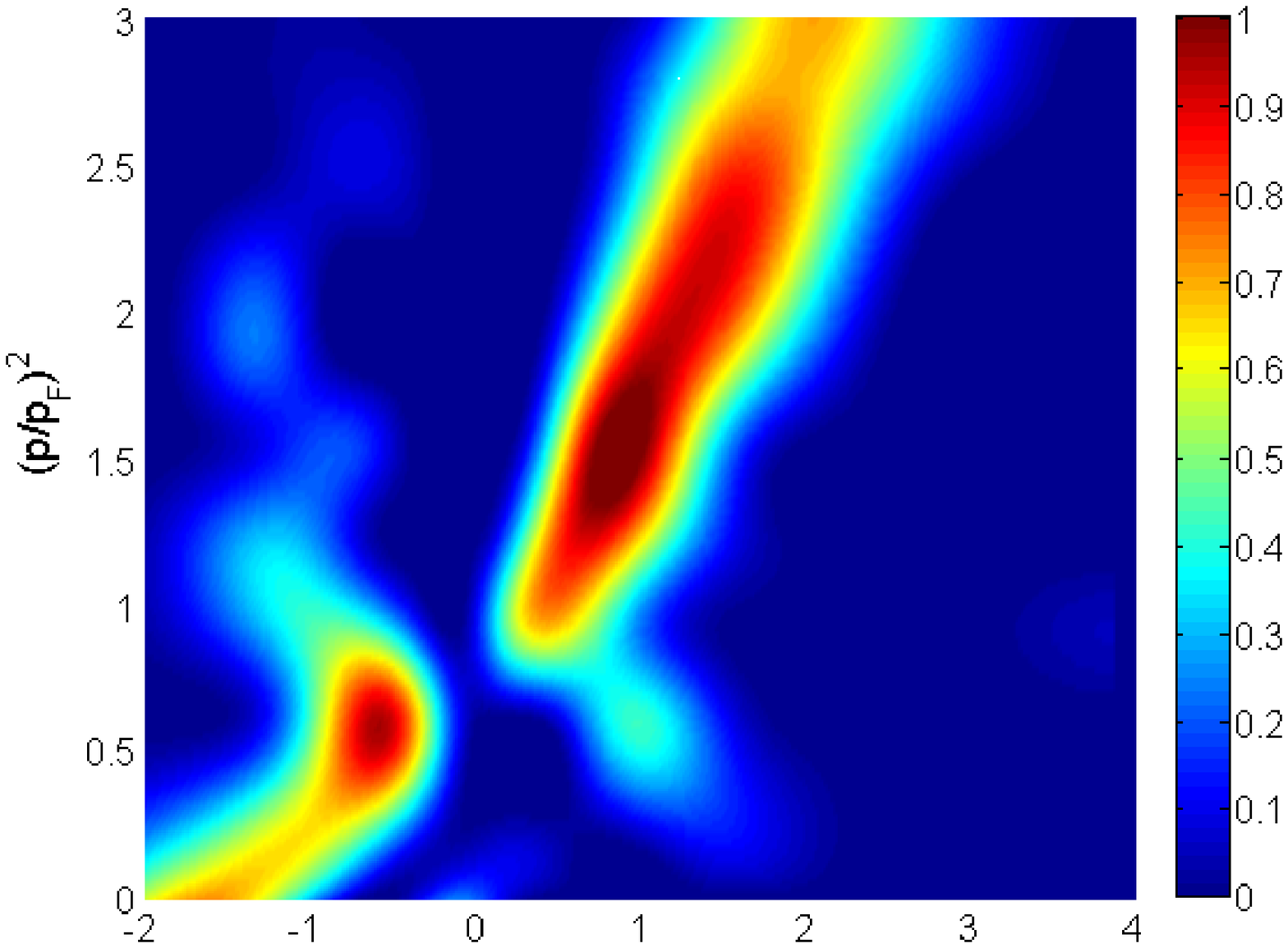}
\includegraphics[width=7.5cm]{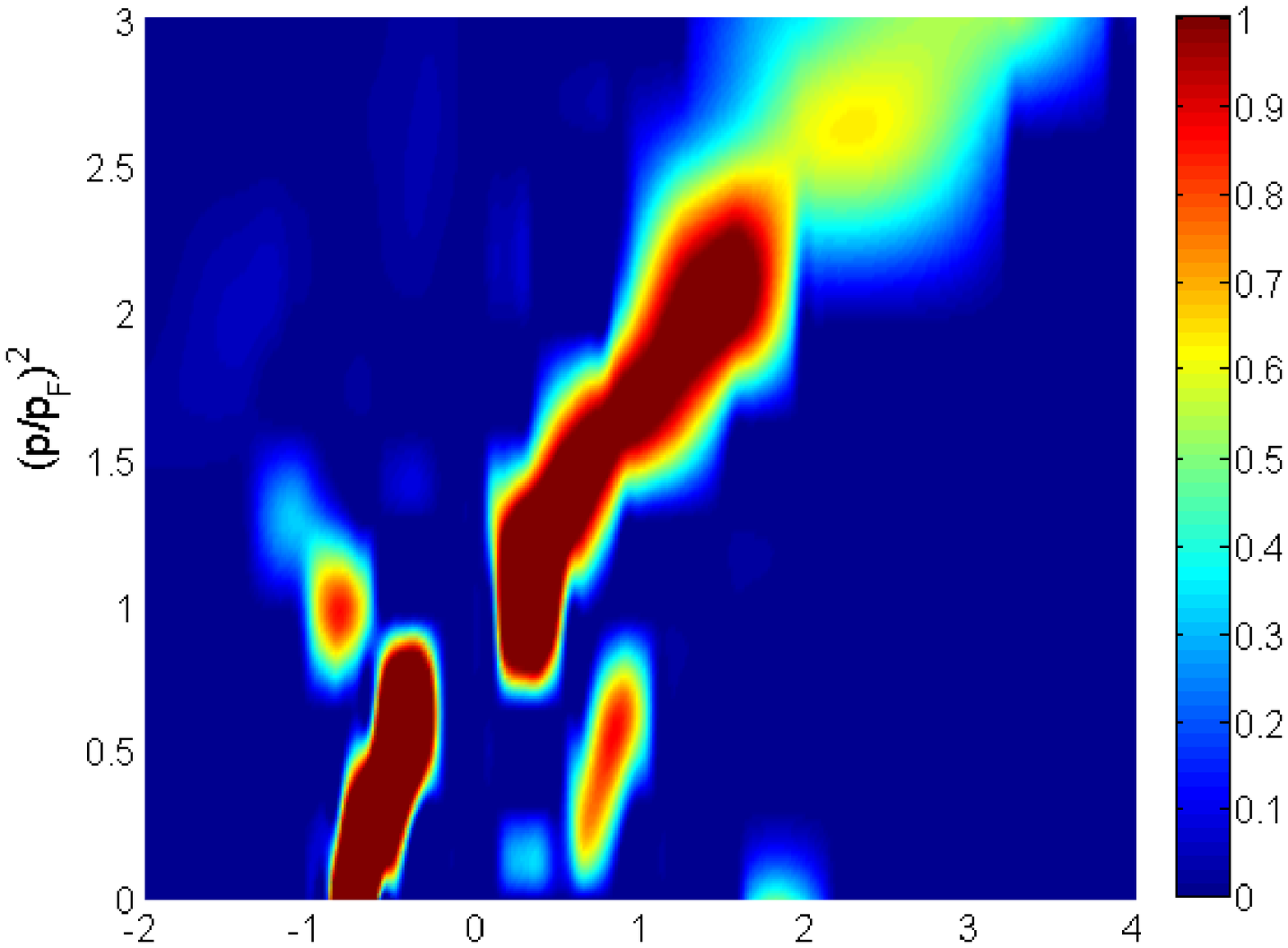}
\includegraphics[width=7.5cm]{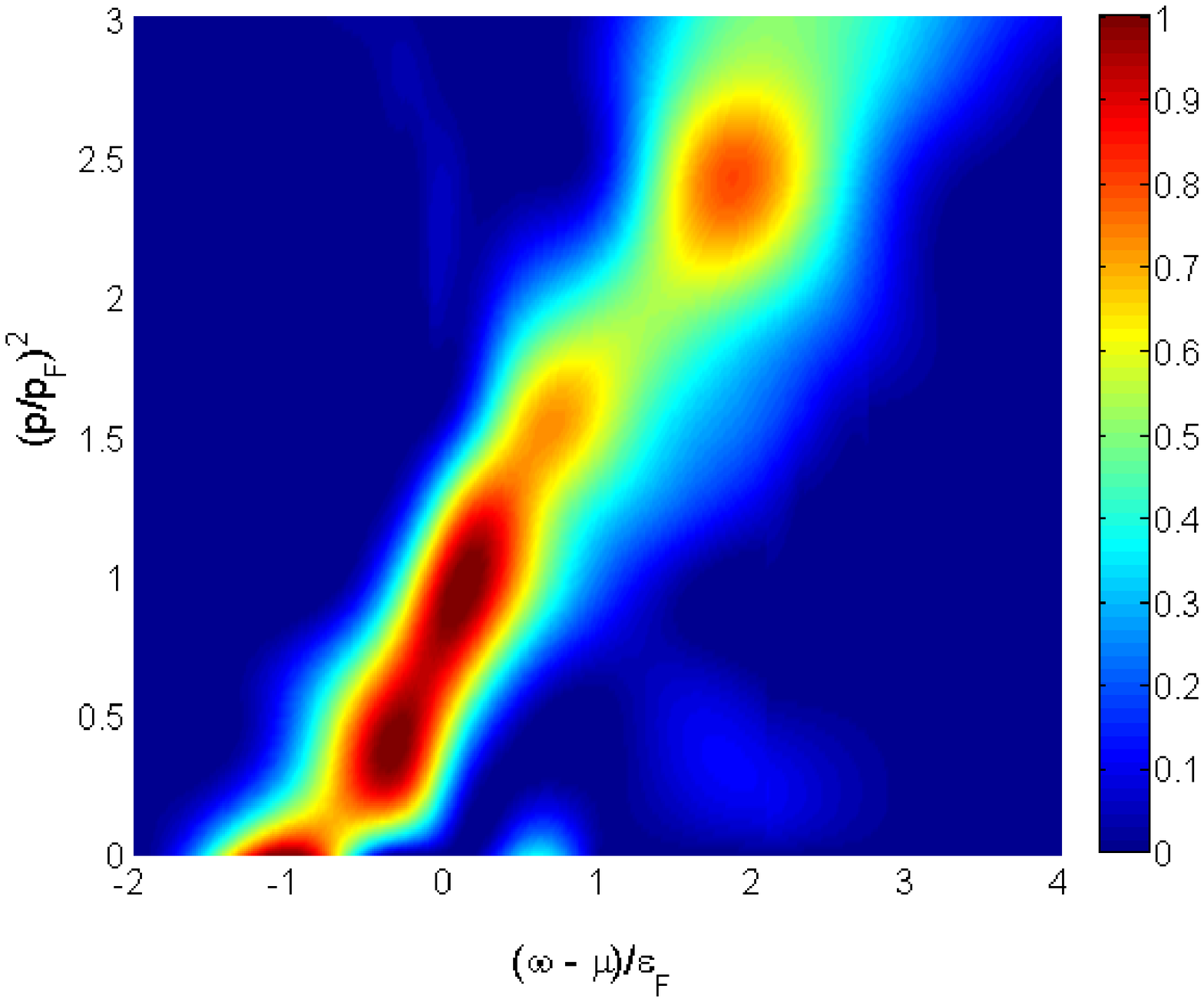}
\caption{ (Color online)
Spectral weight function $A(\bm{p},\omega)$ for three temperatures:
$T=0.15\varepsilon_F\approx T_{c}$ (upper panel),
$T=0.18\varepsilon_F$
(middle panel) and $T=0.20\varepsilon_{F}$ (lower panel).
The presence of a gap in clearly seen in the upper two panels.
\label{fig:A} }
\end{figure}
A sample of calculated spectral weight functions at unitarity are shown in
Fig. \ref{fig:A}. In order to characterize the quasiparticle excitation
spectrum
we have associated with the maximum of $A(\bm{p},\omega)$ the quasiparticle
energy $E(\bm{p})$:
\beq
E(\bm{p}) = \pm \sqrt{ \left ( \frac{p^2}{2m^*}+U-\mu \right )^2+\Delta^2},
\label{eq:ek}
\eeq
where $m^{*}$ is an effective mass, the potential $U$ and
the ``pairing'' gap $\Delta$ depend on temperature, and $\mu$ is an
input parameter. In Fig. \ref{fig:eqp} we compare the spectrum of
elementary fermionic excitations evaluated in Ref. \cite{cr}, with the one
extracted by us from our lowest temperature spectral weight function. Such
comparisons are legitimate because the temperature dependence of various
quantities at $T\le T_c$ is relatively weak, see Ref. \cite{bdm} and the
results below. The agreement between the $T=0$ Monte Carlo
results and the low $T$-limit of our finite-$T$ unrestricted QMC
data validates those results. One should keep in mind
that the effective range corrections are noticeable, since
$r_0=4/\pi\Lambda\approx 0.4$ (in lattice units), where $\Lambda$ is the
cutoff in momentum \cite{bdm}.

\begin{figure}[htb]
\includegraphics[width=7.5cm]{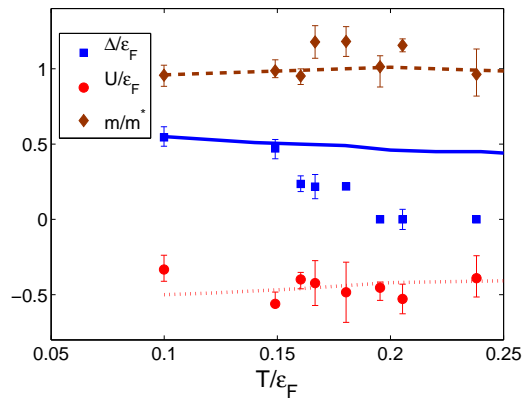}
\caption{ (Color online)
The single-particle parameters extracted from the spectral weight function
at unitarity.
The dashed, solid and dotted lines represent the quantities: $m/m^*$,
$\Delta/\varepsilon_{F}$,
$U/\varepsilon_{F}$, respectively, extracted using the assumption of
independent quasiparticle model.
\label{fig:pmu}}
\end{figure}

The extracted value of the single-particle potential $U$ (see
Fig. \ref{fig:pmu}) shows essentially no temperature dependence in the
range investigated by us. (Simulations at higher temperatures are not
reliable with lattice sizes we considered here, see also Ref. \cite{bdm}.)
The
values of $U$ and $\alpha$ are very close to the values determined
in Ref. \cite{slda} at $T=0$ and show almost no temperature dependence.

A surprising feature of our results can be seen if we assume that the
system is composed
of independent quasiparticles with BCS-like dispersion relation, in which
case the susceptibility can be easily evaluated:
\beq
\chi (\bm{p})=-\int_0^\beta d\tau {\cal G}({\bm p}, \tau)=
\frac{1}{E(\bm{p})}\frac{e^{\beta E(\bm{p})}-1}{e^{\beta E(\bm{p})}+1}.
\label{eq:Ep}
\eeq
From the calculated one-body propagator, using Eqs.
(\ref{eq:Gp},\ref{eq:Ep}), one can
extract the spectrum of the elementary fermionic excitations at
finite temperature, which turns out somewhat unexpectedly
to be accurately parameterized by Eq. (\ref{eq:ek}).
Extracted parameters are plotted in Fig. \ref{fig:pmu} with lines.
While the agreement between the mean-field potential $U$ and effective mass
$m^{*}$ obtained using the two procedures is almost perfect at all
temperatures, the
pairing gap is reproduced satisfactorily only up to $T\le T_c$.

Another notable feature of our results is that both methods admit the
``gapped" spectral function
above the critical temperature, routinely referred to as the pseudogap~\cite{huefner}.
Various aspects and the physics of a pseudogap in a Fermi
gas in the unitary regime have been advocated and discussed for a number of
years
by several groups \cite{levin}. It is however notable that the pseudogap has
not entered the
mainstream of research in this field and the physics of the pseudogap is
barely covered in the recent reviews \cite{reviews}, which reflects a rather
widespread opinion in the cold atom community that the gap should vanish at
$T_c$. There have been several experimental attempts to extract the pairing
gap in ultracold dilute Fermi gases \cite{rf_exp} and a theoretical
explanation
of these spectra was given in Refs. \cite{rf_bcs}. However, it was later
shown in
Refs.~\cite{rf_th} that these initial interpretations of the rf-spectra as
revealing the
pairing gap were in error, as strong final state interaction effects had
been neglected. In particular the difficulties of determining the pairing gap
using rf-spectroscopy were discussed in Ref. \cite{stoof}.
In Ref. \cite{bruun} the use of Bragg spectroscopy is advocated instead, in
order
to measure the onset of superfluidity as well as the appearance of a
pseudogap.
A recent theoretical calculation of the spectral function, based on summing
all the particle-particle ring diagrams \cite{haussmann}, does not reveal
signs of a pseudogap. On the other hand in Refs. \cite{jin,swinburne} there are
experimental indications that a pseudogap exists in a unitary gas.

Our calculations show that the spectral function reveals the presence of a
gap
in the spectrum up to about $T^*\approx 0.20\varepsilon_{F}$. This result is
reproduced by both the maximum entropy and singular value decomposition
methods,
which indicate that the spectral function possesses a two peak structure
around the Fermi level at temperatures above $T_{c}$. In the maximum entropy
method this result
is stable with respect to variation of the assumed model ${\cal M}$
and for $\sigma^{2}\alpha \le 0.3$, see Eq. (\ref{eq:method}).
In range of temperatures $0.18\dots 0.25\varepsilon_{F}$ the singular value
decomposition method
can reproduce both two- and one-peak structures depending on the details of
the method, due to
finite size of the statistical errors in
the imaginary time Green's function. In our case this
resolution is at the level of $\Delta\approx 0.2\dots 0.3\varepsilon_{F}$,
which means that the gap above $T_{c}$ is barely visible by the singular
value decomposition method.
We note that $T^*$ is the temperature at which, not surprisingly in
hindsight, the caloric curve $E(T)$ has a shoulder \cite{bdm}, which in Ref.
\cite{bcs-bec} we called $T_0$.

We thank G.F. Bertsch and M.M. Forbes for discussions and the grant support
from
DOE DE-FG02-97ER41014, DE-FC02-07ER41457, and from the Polish Ministry
of Science No. N N202 328234. Some calculations were performed at ICM of
Warsaw University.

$^*$ Present address: Department of Physics, The Ohio State University,
Columbus, OH 43210-1117, USA.


\end{document}